\documentclass[twocolumn,preprintnumbers, superscriptaddress, reprint, longbibliography, prb]{revtex4-1}
\usepackage{amsmath}
\usepackage{stmaryrd}
\usepackage{txfonts}
\usepackage{amssymb}
\usepackage{mathrsfs}
\usepackage{graphicx}
\usepackage{dcolumn}
\usepackage{bm}
\usepackage{braket}
\usepackage{epsfig}
\usepackage{color}
\usepackage{ulem}
\usepackage{bibunits}

\usepackage[colorlinks=true, linkcolor=blue, citecolor=blue, urlcolor=blue]{hyperref}
\begin{document}
\preprint{\href{https://doi.org/10.1103/PhysRevB.99.140408}{S.-Z. Lin, J.-X. Zhu and A. Saxena, Phys. Rev. B {\bf 99}, 140408(R) (2019).}}

\title{Kelvin modes of a skyrmion line in chiral magnets and the associated magnon transport}
\author{Shi-Zeng Lin}
\email{szl@lanl.gov}
\affiliation{Theoretical Division, T-4 and CNLS, Los Alamos National Laboratory, Los Alamos, New Mexico 87545, USA}
\author{Jian-Xin Zhu}
\affiliation{Theoretical Division, T-4 and CNLS, Los Alamos National Laboratory, Los Alamos, New Mexico 87545, USA}
\affiliation{Center for Integrated Nanotechnologies, Los Alamos National Laboratory, Los Alamos, New Mexico 87545, USA}
\author{Avadh Saxena}
\affiliation{Theoretical Division, T-4 and CNLS, Los Alamos National Laboratory, Los Alamos, New Mexico 87545, USA}

\begin{abstract}

Magnetic skyrmions in bulk crystals are line-like topologically protected spin textures. They allow for the propagation of magnons along the skyrmion line but are localized inside the skyrmion line. Analogous to the vortex line, these propagating modes are the Kelvin modes of a skyrmion line. In crystals without an inversion center, it is known that the magnon dispersion in the ferromagnetic state is asymmetric in the wavevector. It is natural to expect that the dispersion of the Kelvin modes is also asymmetric with respect to the wavevector. We study the Kelvin modes of a skyrmion line in the ferromagnetic background. In contrast, we find that the lowest Kelvin mode is symmetric in the wavevector in the low energy region despite the inversion symmetry breaking. Other Kelvin modes below the magnon continuum are asymmetric, and most of them have a positive group velocity. Our results suggest that a skyrmion line can function as a one-way waveguide for magnons.          

\end{abstract}

\date{\today}
\maketitle

Lord Kelvin calculated stable propagating wave modes along a straight vortex tube of uniform vorticity in a classical fluid  about 130 years ago. \cite{Kelvin1880} These modes were later called Kelvin modes. The Kelvin modes in quantized vortex lines were subsequently studied \cite{Pitaevskii1961,PhysRev.162.143} and observed experimentally \cite{RevModPhys.59.87} in quantum superfluids. The spectrum of the Kelvin modes in the long wavelength limit is $\hbar\omega=\frac{\hbar^2 k^2}{2m} \ln(1/k\xi)$, where $m$ is the mass of a particle in the superfluid, $k$ is the momentum, and $\xi$ is the healing length.  

Recently, a vortex-line like topological spin texture, known as skyrmion, has been observed in magnets by experiments.\cite{Muhlbauer2009,Yu2010a} A large family of skyrmion-hosting materials has been identified. A skyrmion has several intrinsic properties, such as polarization, vorticity and helicity, and these properties are determined by the symmetries of the material and/or external magnetic fields. For skyrmions in materials without inversion symmetry, the Dzyaloshinskii-Moriya interaction \cite{Dzyaloshinsky1958,Moriya60,Moriya60b} (DMI) is responsible for the stabilization of a skyrmion lattice. \cite{Bogdanov89} Skyrmions can also exist in systems with inversion symmetry, where competing magnetic interactions stabilize the skyrmions. \cite{PhysRevLett.108.017206,leonov_multiply_2015,PhysRevB.93.064430,PhysRevB.93.184413,hirschberger_skyrmion_2018,kurumaji_skyrmion_2018} In thin films, skyrmions appear as disk-like excitations and in bulk materials, skyrmions are line-like excitations. Skyrmions can be manipulated by various external drives, such as electric current, electric field, thermal gradient, etc. \cite{Jonietz2010,Yu2012,Schulz2012,White2012,PhysRevLett.113.107203,Kong2013,Lin2014PRL,Mochizuki2014} Remarkably, skyrmions can be driven into motion by a small current density of the order of $10^6\ \mathrm{A/m^2}$, which is $5$ to $6$ orders of magnitude smaller than that for magnetic domain walls.~\cite{Jonietz2010,Yu2012,Schulz2012} For their superior properties including compact size, high mobility and stability, skyrmions have attracted tremendous attention recently and are deemed as promising candidates for applications in the next generation spintronic devices. \cite{Fert2013,nagaosa_topological_2013}   

\begin{figure}[t]
  \begin{center}
  \includegraphics[width=\columnwidth]{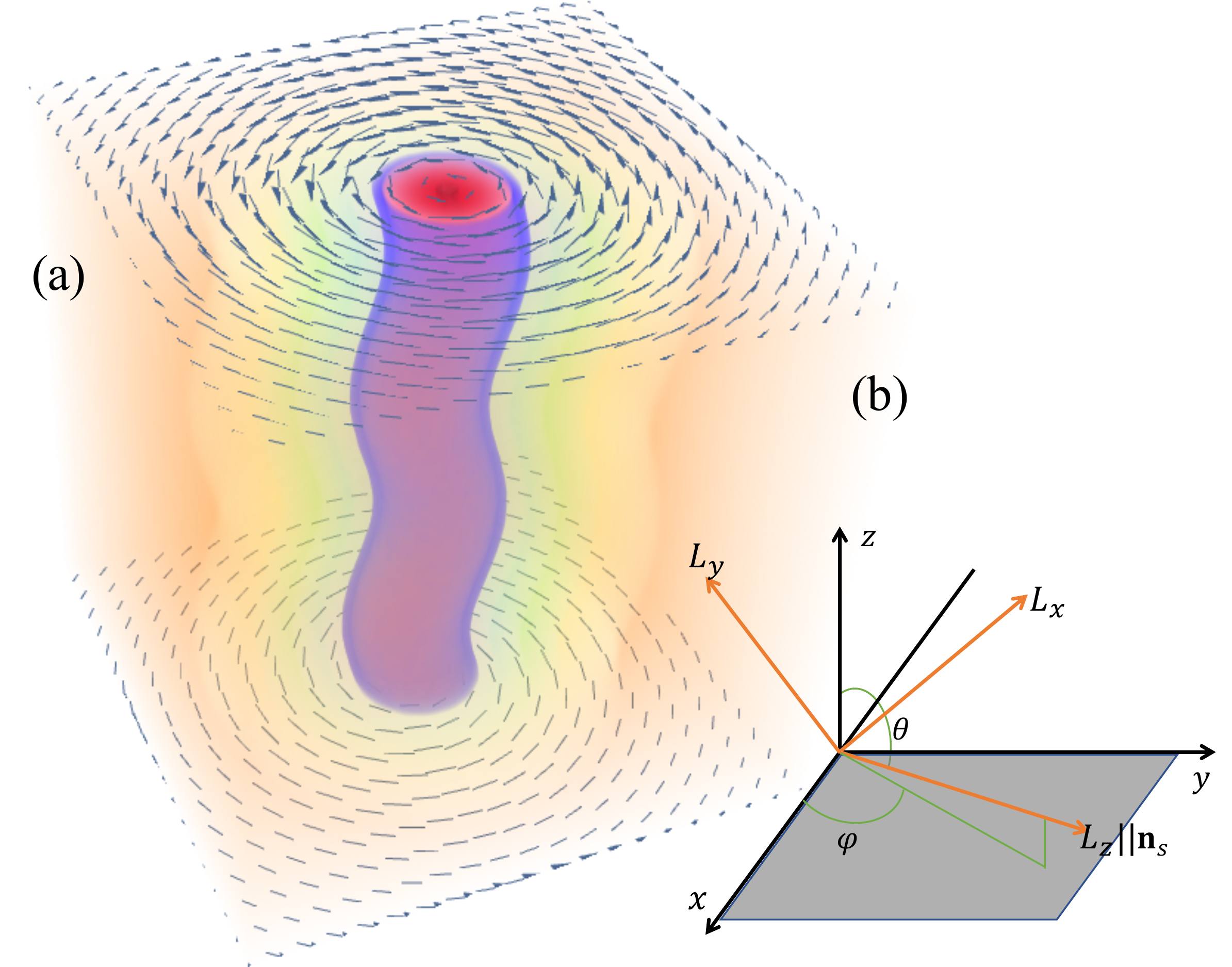}
  \end{center}
\caption{(a) Schematic view of the Kelvin modes of a skyrmion line. (b) Definition of the local spin coordinate $\mathbf{L}$ and its relation to the spin $\mathbf{n}$ in the lab frame.} 
  \label{f1}
\end{figure}

A skyrmion line can also support propagating Kelvin modes inside the line, see Fig.~\ref{f1} (a). In centrosymmetric systems, the Kelvin mode is symmetric in the propagating wavevector. \cite{PhysRevD.90.025010} In systems without inversion symmetry, the question is whether the dispersion of the Kelvin modes is asymmetric with respect to the wavevector. If the answer is positive, this would imply a one-way propagation of magnons inside a skyrmion line, and therefore the skyrmion line can work as a one-way magnon waveguide.  This is the question we will address in this work. 

First, we consider the magnon dispersion in the fully spin polarized state. We begin with a phenomenological description of the magnetization in a chiral magnet. The Hamiltonian of the system in terms of the magnetization field $\mathbf{n}(\mathbf{r})$ with $|\mathbf{n}|=1$ is \cite{bak_theory_1980,Bogdanov89}
\begin{equation}\label{eq1}
\mathcal{H} =\int dr^3\left[ \frac{J}{2} \sum_{\mu  = x,y} {\left( {{\partial _\mu }{\bf{n}}} \right)^2} + D{\bf{n}}\cdot\nabla  \times {\bf{n}} -B_z n_z-\frac{A}{2} n_z^2\right],
\end{equation}
which successfully captures many experimental observations in chiral magnets. Here $J$ is the exchange interaction, $D$ is the DMI \cite{Dzyaloshinsky1958,Moriya60,Moriya60b} and $B_z$ is the external magnetic field. We have introduced an easy axis anisotropy $A>0$. For B20 compounds with cubic symmetry, this term is not allowed, \cite{bak_theory_1980} but this anisotropy can be generated by uniaxial stress. It can also exist in other crystals with a layered structure. We have neglected the weak dipolar interaction. Note that the skyrmion size is much bigger than the spin lattice constant, and this justifies the continuum approximation in Eq. \eqref{eq1}. For the field value above the saturation field, a ferromagnetic state is stabilized, where $\mathbf{n}=\hat{z}$ with $\hat{z}$ being a unit vector in the $z$ direction. \cite{Butenko2010,leonov_properties_2016} The dynamics of $\mathbf{n}$ is determined by the Berry phase contribution to the action $S_B=\frac{S\hbar}{2a^3}\int dr^3 dt \partial_t \varphi  ( \cos\theta  +1)$, where $S$ is the total spin of the ion (in the material) and $a$ is the crystal lattice parameter. Here $\varphi$ and $\theta$ are the spherical angles of $\mathbf{n}$, i.e. $\mathbf{n}=(\sin\theta\cos\varphi,\  \sin\theta\sin\varphi,\  \cos\theta)$. The magnon dispersion is 
\begin{equation}\label{eq2}
    \frac{S\hbar}{a^3}\omega_{\mathrm{FM}}=J \mathbf{k}^2+2D k_z+B_z+A.
\end{equation}
The magnon dispersion is asymmetric with respect to $k_z$ consistent with the inversion symmetry breaking. This asymmetric magnon dispersion has been observed in experiments, \cite{PhysRevB.97.224403,PhysRevLett.120.037203} and provides a useful way to determine the strength of the DMI. Here the ferromagnetic (FM) state is stable for the field above $B_{c}=D^2/J-A$. The asymmetry gives rise to unconventional magnon propagation such as the modified Snell's law. \cite{PhysRevB.94.140410} Note that the asymmetry only appears in $k_z$ along the field direction, while the dispersion with respect to $k_x$ and $k_y$ remain symmetric. Therefore for a thin film with a normal magnetic field, the magnon dispersion is symmetric with respect to the in-plane wave vectors, even though the inversion symmetry is broken.  

A skyrmion line can exist as a metastable state in the background of the FM state. The skyrmion line provides a centrosymmetric potential for magnon excitations, and it allows for the existence of localized magnons. This was calculated in thin films. \cite{PhysRevB.90.094423,Lin_internal_2014} In clean systems, the magnon modes can be labeled by angular momentum $m$ and wavevector $k_z$. In the following discussion, we call these modes the Kelvin modes with quantum numbers $m$ and $k_z$. Let us consider the lowest mode associated with the translation of the skyrmion line. The translation of the whole straight skyrmion line does not cost any energy and it is a Goldstone mode of the system. The bending of the skyrmion line costs energy and results in the dispersion of the corresponding Kelvin mode. The displacement of a rigid skyrmion line can be described by $\mathbf{n}_s[\mathbf{r}-\mathbf{u}(z)]$, where $\mathbf{u}(z)=[u_x(z),\ u_y(z)]$ is the displacement vector. The $z$ independent displacement of a skyrmion line does not cost energy, i.e. $\mathcal{H}[\mathbf{n}_s(\mathbf{r}-\mathbf{u}_0)]=\mathcal{H}[\mathbf{n}_s(\mathbf{r})]$. For a long wavelength distortion, the energy functional can be expanded in the basis of $\partial_z\mathbf{u}$, $\mathcal{F}(\mathbf{u})\propto \int dz (\partial_z\mathbf{u})^2+\cdots$. The first order term $\partial_z\mathbf{u}$ appears as a surface term upon integration. It vanishes when the two ends of the skyrmion line are fixed.

The energy cost to distort a skyrmion line can also be obtained directly from Eq. \eqref{eq1} and is 
\begin{equation}
    E_z=\frac{1}{2} J (\partial_z\mathbf{n})^2 +D\left(n_y\partial_zn_x-n_x\partial_z n_y\right).
\end{equation}
The contribution from the DMI vanishes as obtained by straightforward calculations. Therefore the distorted skyrmion line has an energy cost 
\begin{equation}
E_z=\frac{1}{2} \eta  J \int dz \left[(\partial_z u_x)^2+(\partial_z u_y)^2\right],    
\end{equation}
with $\eta\equiv\int dr^2(\partial_x \mathbf{n}_s)=\int dr^2(\partial_y \mathbf{n}_s)$. The stiffness of the skyrmion line is $\eta J$ and is independent of the DMI, which is consistent with the fact that $J$ is the largest energy scale of the problem.    

In terms of $\mathbf{u}(z)$, the Berrry phase part of the action becomes
\begin{equation}
S_B=S_B(\mathbf{u}=0)+\frac{S \hbar  \pi }{a^3}\left(u_x {\partial_t u_y}-u_y {\partial_t u_x}\right).    
\end{equation}
The total action associated with the distortion of a skyrmion line is $S_T=S_B-E_z$. The equation of motion for $\mathbf{u}(z)$ is
\begin{eqnarray}
\frac{2\pi\hbar S}{a^3} \partial_t u_y-J\eta \partial _z^2u_x=0,\\
-\frac{2\pi\hbar S}{a^3} \partial_t u_x-J\eta \partial _z^2u_y=0.
\end{eqnarray}
Therefore the dispersion of this Kelvin mode is (as will be shown below, this Kelvin mode has $m=\pm 1$)
\begin{equation}\label{eqKM8}
    \omega =\frac{a^3 \eta  J }{2\pi\hbar S}k_z^2.
\end{equation}
This Kelvin mode is symmetric with respect to $k_z$. This is different from the magnon mode in the ferromagnetic state, where the magnon dispersion is asymmetric due to the presence of DMI. The mode is gapless. A gap exists when there is a local pinning potential or geometric confinement in a small system. The gap can be introduced into the action $S_T$ by adding a mass term $M \mathbf{u}^2/2$ for a straight skyrmion line. Nevertheless, the dispersion of other Kelvin modes is asymmetric with respect to $k_z$ as will be shown below.

To go beyond the analysis of the skyrmion displacement field, we calculate the magnon spectrum in the presence of a straight skyrmion line embedded in the ferromagnetic background. For a disk-like skyrmion in thin films, the spectrum was calculated in in Refs. \onlinecite{PhysRevB.90.094423,Lin_internal_2014}. Here we extend the method used in Ref. \onlinecite{PhysRevB.90.094423} to three dimensions. First, we find the stationary solution of a straight skyrmion line. The symmetry of the problem allows us to use cylindrical coordinates $\mathbf{r}=(r,\ \phi,\ z)$. The skyrmion line solution in Eq. \eqref{eq1} has the form $\varphi=\phi+\pi/2$ (skyrmion helicity is $\pi/2$ determined by the DMI) and $\theta(r)$ with $\theta$ changing from $\theta=-\pi$ at the skyrmion center $r=0$ to $\theta=0$ at $r=\infty$. We obtain the equation for $\theta(r)$ by minimizing $\mathcal{H}$
\begin{align}
    \frac{J}{{2r}}\sin \left( {2\theta } \right) + D\cos \left( {2\theta } \right) + {B_z}r\sin\theta  + \frac{A}{2}r\sin \left( {2\theta } \right)
    \nonumber\\ 
    - \left( {J{\partial _r}\theta  + D} \right) - r J\partial _r^2\theta  = 0,
\end{align}
from which $\theta(r)$ can be found numerically.

We then introduce a local coordinate system with the local $z$ axis along the spin direction $\mathbf{n}_s(\mathbf{r})$. The spin representation in the lab coordinate and the local coordinate is sketched in Fig.~\ref{f1} (b). The local coordinate is obtained by the subsequent rotation operations in the lab frame: rotation along the $z$ axis by $\phi_0=\pi/2$, rotation along the $y$ axis by $\theta$ and rotation along the $z$ axis by $\varphi$. Then the spin in the lab frame $\mathbf{n}$ can be obtained from the local coordinate $\mathbf{L}=(L_X,\ L_Y,\ L_Z)$ according to $\mathbf{n}=\hat{O}\mathbf{L}$, with
\begin{equation}\label{eq4}
\hat{O}= \left(
\begin{array}{ccc}
 -\text{sin$\varphi $} & -\text{cos$\varphi $} \text{cos$\theta $} & \text{cos$\varphi $} \text{sin$\theta $} \\
 \text{cos$\varphi $} & -\text{sin$\varphi $} \text{cos$\theta $} & \text{sin$\varphi $} \text{sin$\theta $} \\
 0 & \text{sin$\theta $} & \text{cos$\theta $} \\
\end{array}
\right).
\end{equation}
The small deviations $\mathbf{L}$ from the skyrmion line solution $\bar{L}_{X}=\bar{L}_Y=0$ and $\bar{L}_Z=1$ are described by the complex magnon fields
\begin{equation}\label{eq5}
\psi=\frac{L_X+i L_Y}{\sqrt{2}}, \ \ \ \psi^*=\frac{L_X-i L_Y}{\sqrt{2}},
\end{equation}
and $L_Z=1-\psi\psi^*$ with $|\psi|\ll 1$. Expanding the Hamiltonian to second order in $\psi$, we obtain
\begin{equation}\label{eq6}
\mathcal{H}_\psi=\frac{1}{2}\hat{\psi}^\dagger{H}_\psi\hat{\psi}, ~~~ \hat{\psi}^\dagger=(\psi^*,\ \  \psi),
\end{equation}
\begin{align}\label{eq7}
{H}_\psi=(-J \nabla^2+V_0)\sigma_0+V_1\sigma_x\nonumber\\
-2\sigma_z\left[\left(J\frac{\cos\theta}{r^2}-D\frac{\sin\theta}{r}\right)i\partial_\phi-i D\cos\theta\partial_z\right],
\end{align}
with $\sigma_i$ ($i=x,\ y,\ z$) being the Pauli matrices and $\sigma_0$ is the unit matrix. Here
\begin{align}\label{eq9}
{V_0} = J\frac{{1 + 3\cos \left( {2\theta } \right)}}{{4{r^2}}} -D \frac{{3\sin \left( {2\theta } \right)}}{{2r}} + {B_z}\cos\theta  - D {\partial _r}\theta \nonumber\\
- \frac{J}{2}{\left( {{\partial _r}\theta } \right)^2}-\frac{A}{2}\left(1-3\cos^2\theta\right),
\end{align}
\begin{equation}\label{eq10}
{V_1} = J \frac{{{{\sin }^2}\theta }}{{2{r^2}}} + D \frac{{\sin \left( {2\theta } \right)}}{{2r}} - D {\partial _r}\theta  - \frac{J}{2}{\left( {{\partial _r}\theta } \right)^2}+\frac{A}{2}\sin^2\theta.
\end{equation}
The presence of skyrmion gives rise to an emergent magnetic field acting on the magnons. This can be seen explicitly by introducing an effective vector potential
\[
\mathbf{a}=-\hat{\phi } \left(\frac{\cos\theta }{r }-\frac{D \sin\theta}{J}\right)+\frac{ D \cos\theta }{J}\hat{z},
\]
with $\hat{\phi}$ being the unit vector in the $\phi$ direction. Using $\nabla\cdot \mathbf{a}=0$, $H_\psi$ can be written in a compact form
\begin{equation}
H_\psi=J{\left( { - {{i}}\nabla  - {\sigma _3}{\bf{a}}} \right)^2} + {\sigma _0}\left( {{V_0} - J{{\bf{a}}^2}} \right) + {\sigma _1}{V_1}.
\end{equation}
The emergent vector potential $\mathbf{a}$ couples to the magnons and induces a screw scattering of the extended magnons by skyrmions.\cite{PhysRevB.89.064412,PhysRevB.90.094423}

\begin{figure}[t]
  \begin{center}
  \includegraphics[width=\columnwidth]{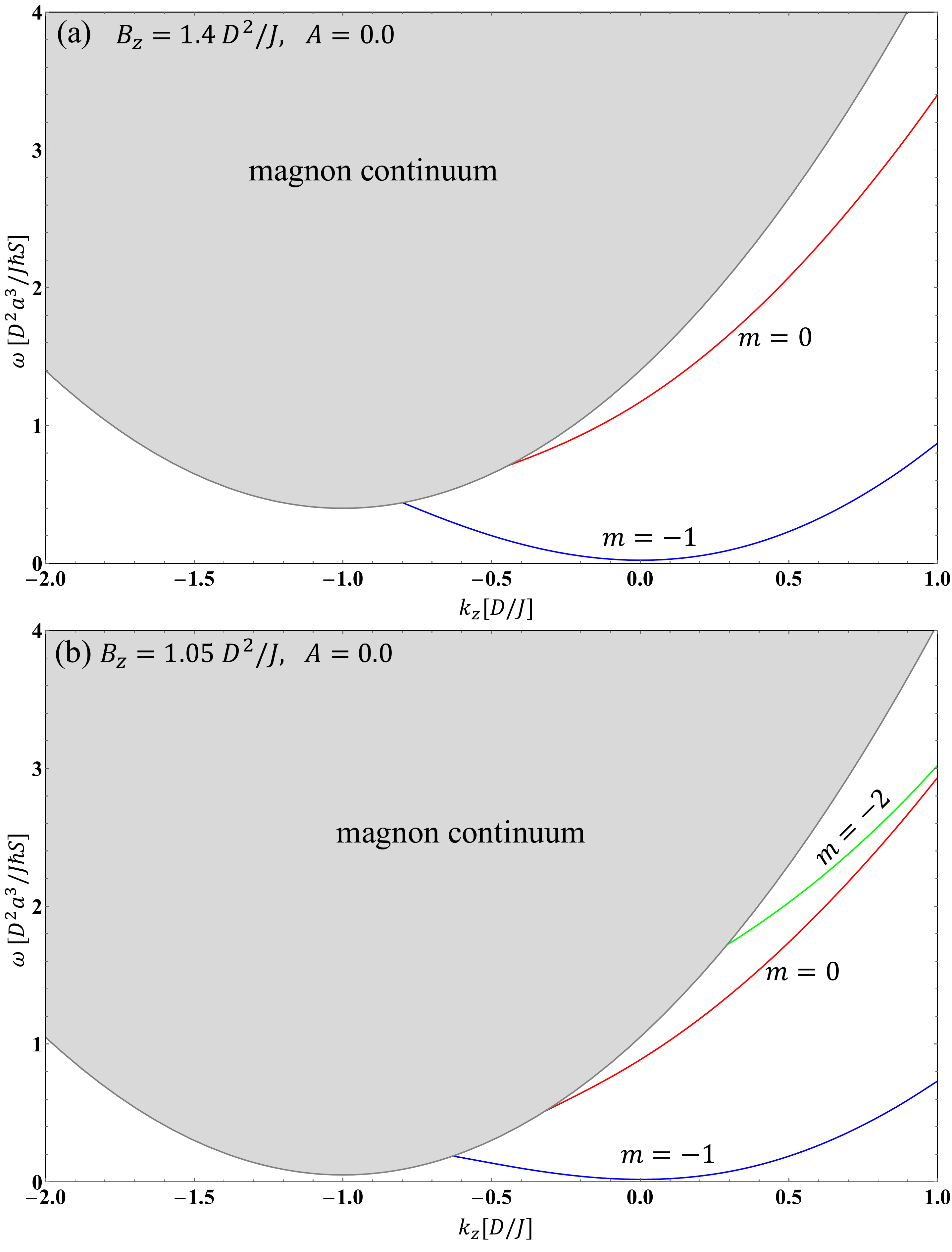}
  \end{center}
\caption{Dispersion of the Kelvin modes with angular momentum $m$. The gray region is the mangon continuum.} 
  \label{f2}
\end{figure}

The eigenmodes are determined by the equation
\begin{equation}\label{eq11}
-i\frac{S\hbar}{a^3}\sigma_z\partial_t\hat{\psi}={H}_\psi\hat{\psi}.
\end{equation}
This equation has the form of the Schr\"{o}dinger equation describing the magnon wave function in a centrosymmetric potential. We can introduce an angular momentum $m$ and wavevector $k_z$ with $\psi=\psi_m(r,t)\exp(i m\phi+i k_z z)$ to label the eigenmodes. The two components of $\hat{\psi}$ are related by complex conjugation because the magnetic moment $\mathbf{n}$ is real. This indicates that the matrix equation, Eq. \eqref{eq11}, is redundant. Indeed ${H}_\psi$ has particle-hole symmetry, ${H}_\psi=\sigma_x K {H}_\psi K \sigma_x$ with $K$ being the complex conjugate operator. This means that if $\exp[i(\omega t+m\phi+k_z z)]\hat{\eta}_m$, with $\hat{\eta}_m^\dagger\equiv (\eta_1^*, \ \ \eta_2^*)$, solves Eq. \eqref{eq11}, then $\exp[-i(\omega t+m\phi+k_z z)]\sigma_x K\hat{\eta}_m$ also solves Eq. \eqref{eq11}. We therefore only take the magnon branch with $\omega\ge 0$. Then $\hat{\psi}$ can be obtained by a linear superposition of the two symmetry-related solutions
\begin{equation}\nonumber
\hat{\psi}_m=b \exp[i(\omega t+m\phi+k_z z)]\hat{\eta}_m+b^*\exp[-i(\omega t+m\phi+k_z z)]\sigma_x K\hat{\eta}_m.
\end{equation}
The two components of $\hat{\psi}_m$ are complex conjugate to each other. Here $\hat{\eta}$ is determined by the eigenvalue problem
 \begin{equation}\label{eq13}
\frac{S\hbar}{a^3}\omega_m\sigma_z\hat{\eta}_m={H}_\psi\hat{\eta}_m.
\end{equation}
When the frequency is much larger than the magnon gap of the FM state, $\frac{S\hbar}{a^3}\omega_{g}=B_z+A-D^2/J$, i.e.  $\omega\gg \omega_g$, the magnon dispersion reduces to that in Eq. \eqref{eq2} and the eigenmodes are $\hat{\eta}_m^\dagger=(1,\ 0)J_m(k r)$. We represent the matrix ${H}_\psi$ using the Bessel function $J_m(k r)$ as an orthogonal basis. \cite{PhysRevB.96.014407} The basis functions are 
\begin{align}\nonumber
|{p_{m,i}}\rangle  = \frac{{\sqrt 2 }}{{{R_c}{J_m}\left( {{k_{m - 1,i}}} \right)}}{J_{m - 1}}\left( {{k_{m - 1,i}}\frac{r}{{{R_c}}}} \right)\exp \left( {{{i}}m\phi+i k_z z } \right)\left( {\begin{array}{*{20}{c}}
1\\
0
\end{array}} \right),
\end{align}
\begin{align}\nonumber
|{h_{m,i}}\rangle  = \frac{{\sqrt 2 }}{{{R_c}{J_{m + 2}}\left( {{k_{m + 1,i}}} \right)}}{J_{m + 1}}\left( {{k_{m + 1,i}}\frac{r}{{{R_c}}}} \right)\exp \left( {{{i}}m\phi+i k_z z  } \right)\left( {\begin{array}{*{20}{c}}
0\\
1
\end{array}} \right),
\end{align}
where we have used the box normalization with $R_c$ being the radius of the box and $k_{m,i}$ is the $i$-th zero of the Bessel function $J_m(kr)$. Then the matrix elements of ${H}_\psi$ are 
\begin{align}
\hat{{H}}_{11;ij}^{(m)}=\langle p_{m,i} | {H}_\psi |{p_{m,j}}\rangle, \ \  \hat{{H}}_{12;ij}^{(m)}=\langle p_{m,i} | {H}_\psi |{h_{m,j}}\rangle, \nonumber\\
\hat{{H}}_{21;ij}^{(m)}=\langle h_{m,i} | {H}_\psi |{p_{m,j}}\rangle, \ \  \hat{{H}}_{22;ij}^{(m)}=\langle h_{m,i} | {H}_\psi |{h_{m,j}}\rangle.
\end{align}
By diagonalizing the matrix $\sigma_z {H}_\psi$, we obtain the eigenfrequencies and eigenmodes. We take $R_c=20$ and truncate the Bessel series at $i_{\mathrm{max}}=20$. 

\begin{figure}[t]
  \begin{center}
  \includegraphics[width=\columnwidth]{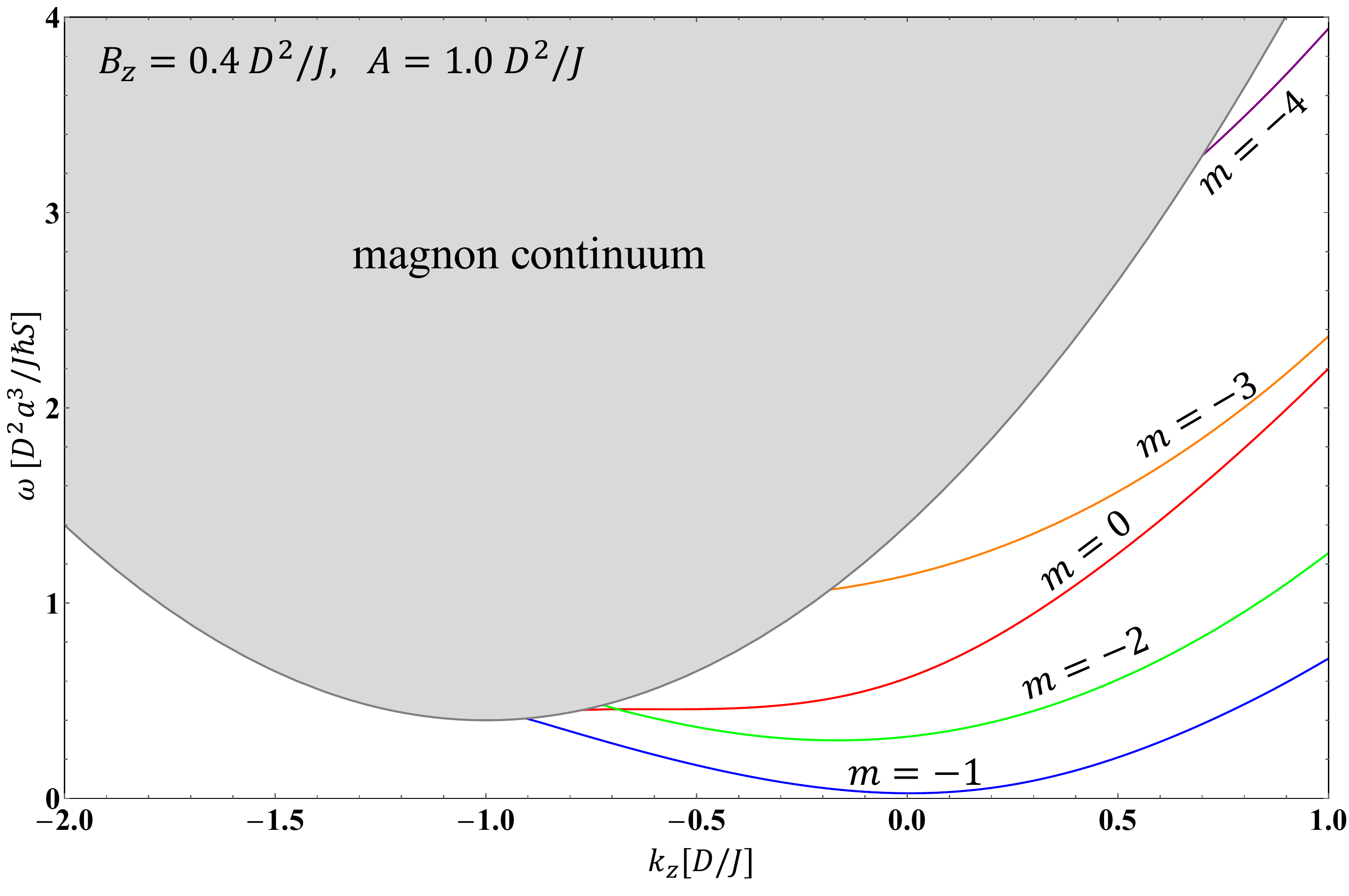}
  \end{center}
\caption{The same as Fig. \ref{f2}, but with an easy axis anisotropy, $A=1.0\ D^2/J$ and $B_z=0.4\ D^2/J$.} 
  \label{f3}
\end{figure}

The calculated dispersion of the Kelvin modes with different $m$ is shown in Fig.~\ref{f2}. There are only two Kelvin modes below the magnon continuum when $B_z=1.4 D^2/J$ in Fig. \ref{f2} (a), and as a consequence, these modes are radially localized inside the skyrmion. The other modes mix with the magnon continuum and can easily decay into the extended magnon modes. The Kelvin mode with $m=-1$ corresponds to the distortion of a rigid skyrmion line discussed above. It is symmetric with respect to $k_z$ in the low energy region consistent with that in Eq. \eqref{eqKM8}. The mode with $m=0$ corresponds to the uniform radial breathing of the skyrmion line. The group velocity $v_{g}=d\omega/dk_z$ is always positive, indicating a one-way propagation of this Kelvin mode. At a lower field, $B=1.05D^2/J$ in Fig. \ref{f2} (b), the Kelvin mode with $m=-2$ also appears below the magnon continuum. In the presence of an easy axis anisotropy, there appear more Kelvin modes below the magnon continuum, see Fig. \ref{f3}. The Kelvin mode with $m=-1$ in Figs. \ref{f2} and \ref{f3} has a very small gap originating from the numerical discretization in the calculations, which breaks the translation symmetry. The left branch of the Kelvin mode with $m=-1$ at high energy merges into the magnon continuum, and therefore is strongly damped. It only allows magnons with a positive group velocity to propagate in this region. Here all the Kelvin modes with $m\neq -1$ are gapped, which guarantee the meta-stability of the skyrmion line in the ferromagnetic background.   

To use the skyrmion line as a one-way magnonic waveguide, it is required to excite the Kelvin mode with $m\neq -1$. This can be achieved by choosing the angular momentum of the source field. Recently, the propagation of magnons both with symmetric and asymmetric dispersion in a skyrmion line in prismatic geometry was demonstrated using micromagnetic simulations. \cite{xing_skyrmion_2019} The propagation of linear magnon wave and nonlinear solitary wave excitations along a skyrmion line in chiral magnets was considered in Ref. \onlinecite{kravchuk_solitary_2019}. The results on the linear magnon wave are consistent with ours. The unidirectional propagation of magnon in the skyrmion line crystal in $\mathrm{Cu_2OSeO_3}$ was investigated both experimentally and theoretically. \cite{seki_propagating_2019}

We have focused on a system with DMI form in Eq. \eqref{eq1}, which can be realized in crystals having $D_n$ or $C_n$ symmetry, \cite{Bogdanov89} for example B20 chiral magnets including FeGe and MnSi. For crystals with $C_{nv}$ or $D_{2d}$ symmetry, such as Mn-Pt-Sn Heusler materials,  \cite{nayak_magnetic_2017} $\mathrm{GaV_4S_8}$, \cite{kezsmarki_neel-type_2015}  $\mathrm{GaV_4Se_8}$,  \cite{PhysRevB.95.180410} and $\mathrm{VOSe_2O_5}$ \cite{PhysRevLett.119.237201} no spatial derivative along the crystal $c$ axis is allowed in the DMI. \cite{PhysRevB.96.214413} In these systems, all the Kelvin modes are symmetric with respect to $k_z$.  

It is possible that skyrmion lines that do not percolate the whole system are stabilized. \cite{Milde2013,yokouchi_current-induced_2018,SZLinMonopole2016,PhysRevB.94.174428,PhysRevB.98.054404} At the ends of the lines, there appear emergent magnetic monopoles or antimonopoles. When both ends of a skyrmion line are terminated by a monopole and an antimonopole, the skyrmion line serves as a magnonic cavity for the Kelvin modes because these localized modes cannot penetrate into the ferromagnetic state. In imperfect systems, the skyrmion line is distorted in order to accommodate the pinning potential. \cite{Blatter94} The pinning opens a gap for the symmetric lowest Kelvin mode. The bent skyrmion line can still guide the Kelvin modes to propagate along the line.

To summarize, we have studied the Kelvin modes of a straight skyrmion line in chiral magnets. There exist several Kelvin modes below the magnon continuum, and these modes are radially localized in the skyrmion line. The Kelvin mode with angular momentum $m=-1$ is symmetric with respect to the wavevector along the skyrmion line in the low energy region. The Kelvin modes with other $m$ are asymmetric. Our results suggest that the skyrmion lines can function as a one-way magnonic waveguide.

\begin{acknowledgments}
SZL would like to thank Congjun Wu and Daniel P. Arovas for motivating the present study. The authors thank Markus Garst for useful discussions and for sharing their results prior to publication. This work was carried out under the auspices of the U.S. DOE NNSA under contract No. 89233218CNA000001 through the LDRD Program and the U.S. DOE Office of Basic Energy Sciences Program E3B5 (SZL and JXZ).
\end{acknowledgments}

\bibliography{references}

\end{document}